\documentclass{pnastwo}

\usepackage{graphicx}

\usepackage{amssymb,amsfonts,amsmath}

\contributor{Submitted to Proceedings
of the National Academy of Sciences of the United States of America}
\url{www.pnas.org/cgi/doi/10.1073/pnas.0709640104}
\copyrightyear{2008}
\issuedate{Issue Date}
\volume{Volume}
\issuenumber{Issue Number}
\begin{document}

%%%%%%%%%%%%%%%%%%%%%%%%%%%%%%

\title{The ``true model'' myth}

\author{Charles T. Perretti\affil{1}{Scripps Institution of Oceanography, University of California, San Diego},
Stephan B. Munch\affil{2}{Fisheries Ecology Division, Southwest Fisheries Science Center},
George Sugihara\affil{1}{}}

\contributor{Submitted to Proceedings of the National Academy of Sciences
of the United States of America}

%% The \maketitle command is necessary to build the title page.
\maketitle

%%%%%%%%%%%%%%%%%%%%%%%%%%%%%%%%%%%%%%%%%%%%%%%%%%%%%%%%%%%%%%%%
\begin{article}

%% When adding keywords, separate each term with a straight line: |
%\keywords{Forecasting | Nonlinear | MCMC | State-space reconstruction}

We have shown that statistical methods widely used in ecology and wildlife management to estimate parameters and forecast population sizes fail when the underlying ecological system is chaotic \cite{perretti2013model}. Although we noted that alternative methods exist \cite{perretti2013model}, their general applicability has not been demonstrated. Hartig \& Dormann \cite{hartig2013does} show that the Pisarenko \& Sornette (2004) method can apply to the logistic model in particular and can provide reasonable parameter estimates. 

Such results satisfy our intuitive expectation that the ``true model'' ought to work the best, provided we know the ``true model'' to begin with. However, one must question whether we ever know the ``true model'' in ecology. Rather, we routinely acknowledge that all models are ultimately wrong while hoping that some models provide a useful approximation. Therefore an extremely relevant issue arising from our original work is the question of how well any explicit model fitting methods improve our ability to forecast with an approximate but ultimately incorrect model.

To simulate this situation, we generated time series using the theta-logistic model ($N_{t+1} = N_{t}r (1 - (N_{t}/K)^\theta)$), and then fit the logistic model using the method employed by Hartig \& Dormann \cite{hartig2013does}. The functional form of the theta-logistic is identical to that of the logistic when a single parameter, $\theta = 1$. We expect therefore that the logistic should be a reasonable approximating model.

Similar to previous work \cite{perretti2013nonparametric}, we find that the model fitting method suggested by Hartig and Dormann \cite{hartig2013does} provides inaccurate predictions and severely biased estimates of the dynamics (Fig. 1). This occurs in the absence of observation error, and despite model-fit diagnostics that suggest convergence to a correct model. In contrast, the model-free method of state-space reconstruction (SSR) avoids any requirement for ad-hoc assumptions as to model structure and provides useful short-term forecasts that can be validated empirically. Moreover, a simple extension of the model-free method allows one to test potential mechanisms and causal drivers \cite{sugihara2012detecting}, thus promoting a better mechanistic understanding without reliance on a mythical ``true model.''

%% == end of paper:

\begin{figure}
\centering
\includegraphics[width=6.5cm]{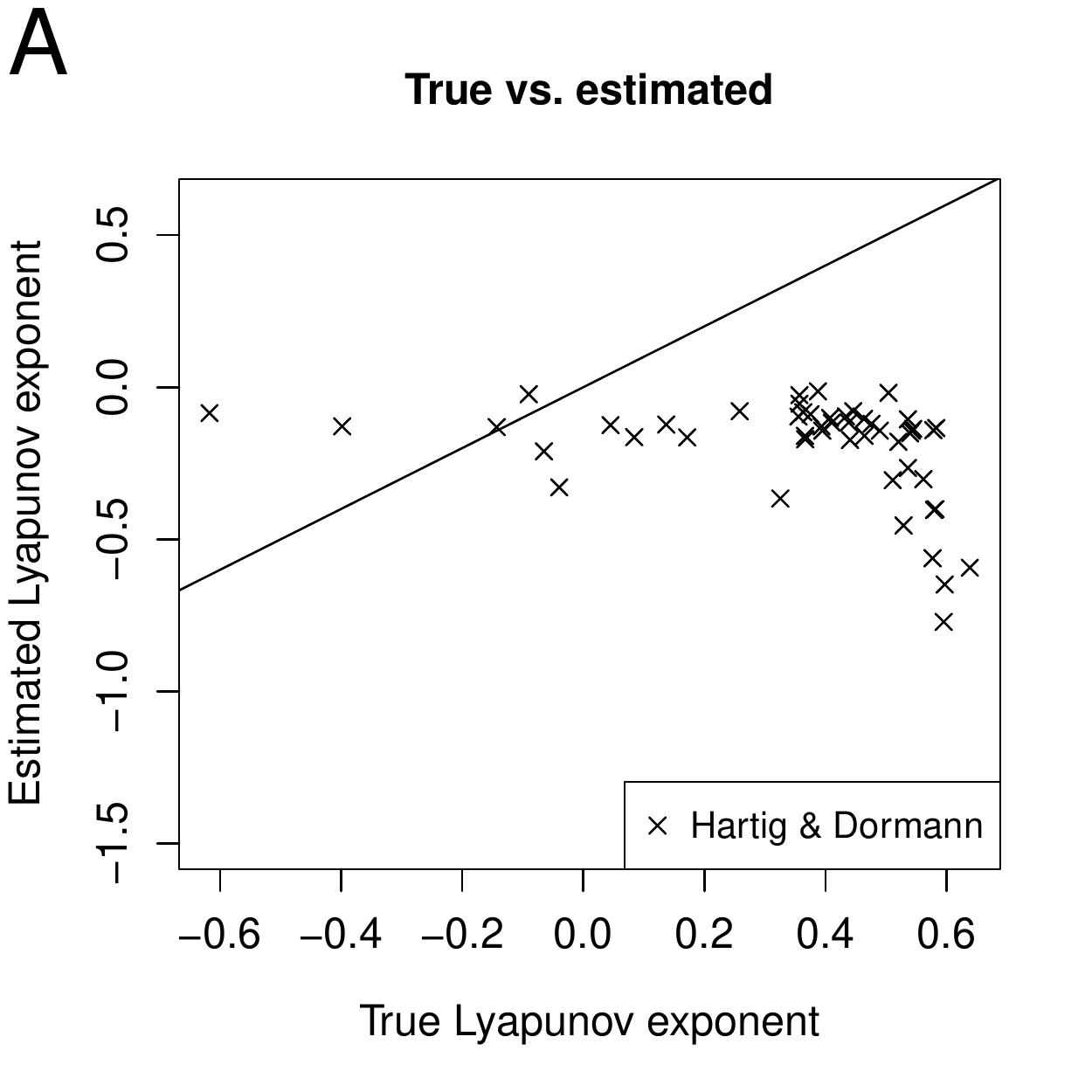}
\includegraphics[width=6.5cm]{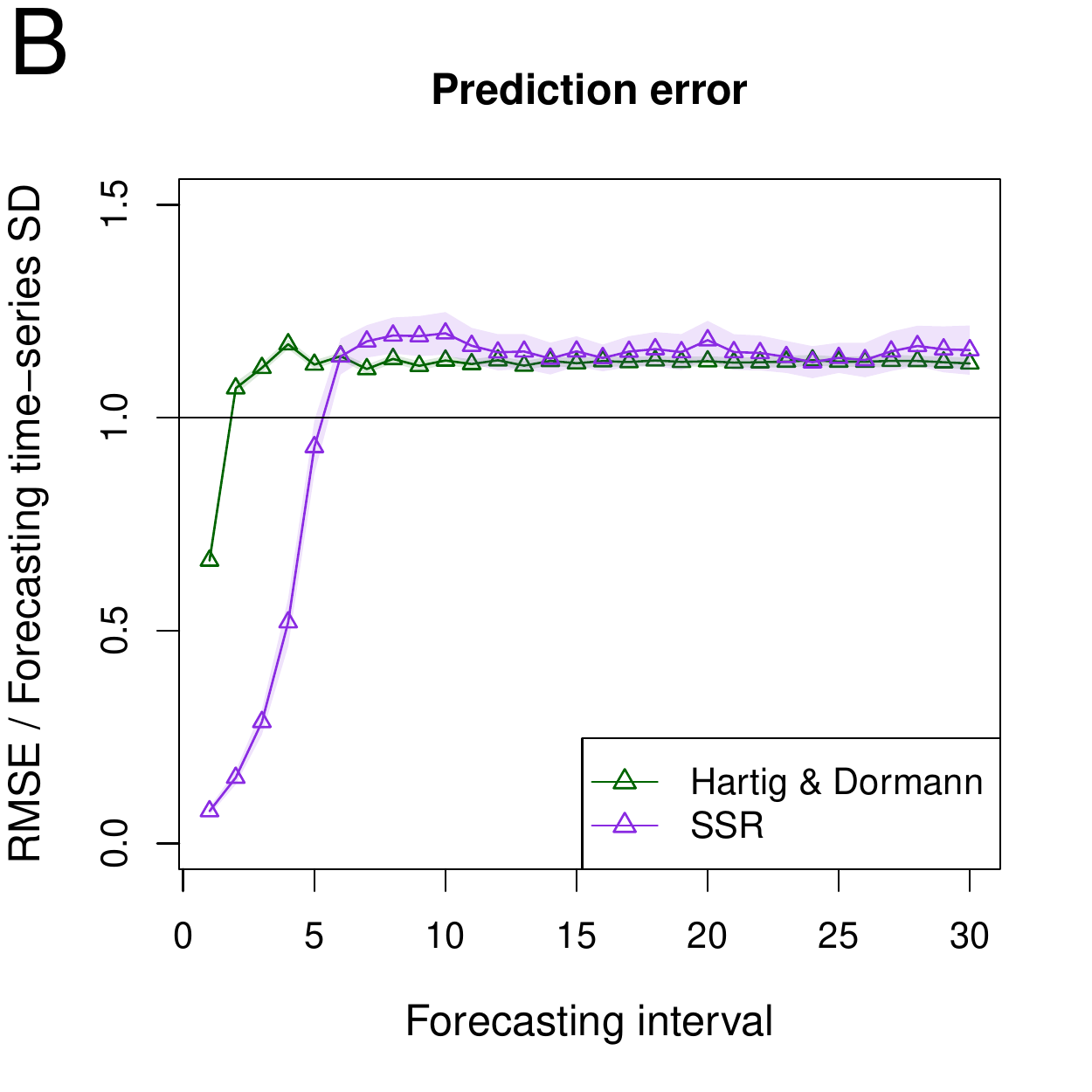}
\caption{Estimated Lyapunov exponent of the logistic model (A), and predictive accuracy (B) when the true model is the theta-logistic model (for $r=1.8$, and $\theta = 4.0$). Lyapunov exponent estimates using the method proposed by Hartig \& Dormann are consistently biased towards stable dynamics (Lyapunov exponent $<$ 0) when the true dynamics were often chaotic (Lyapunov exponent $>$ 0). In addition, the SSR method exhibits substantially lower forecast error than the fitted logistic model (B). Forecast error is represented by standardized RMSE, the shaded interval represents the 95\% confidence interval.} 
\end{figure}

\end{article}

\end{document}